# The optical properties of few-layer InSe

Chaoyu Song, Shenyang Huang, Chong Wang, Jiaming Luo, Hugen Yan*

**Affiliations:** State Key Laboratory of Surface Physics, Key Laboratory of Micro and Nano Photonic Structures (MOE) and Department of Physics, Fudan University, Shanghai 200433, China

Author to whom correspondence should be addressed: E-mail: hgyan@fudan.edu.cn

**Abstract:**

Few-layer InSe draws tremendous research interests owing to the superior electronic and optical properties. It exhibits high carrier mobility up to more than 1000 $cm^2$/Vs at room temperature. The strongly layer-tunable band gap spans a large spectral range from near-infrared to the visible. In this perspective, we systematically review the optical properties of few-layer InSe. Firstly, the intrinsic optical and electronic properties are introduced. Compared to other two-dimensional (2D) materials, the light-matter interaction of few-layer InSe is unusual. The band gap transition is inactive or extremely weak for in-plane polarized light, and the emission light is mainly polarized along the out-of-plane direction. Secondly, we will present several schemes to tune the optical properties of few-layer InSe such as external strain, surface chemical doping and van der Waals (vdW) interfacing. Thirdly, we survey the applications of few-layer InSe in photodetection and heterostructures. Overall, few-layer InSe exhibits great potential not only in fundamental research, but also in electronic and optoelectronic applications.

**I. Introduction**

The isolation of graphene by mechanical exfoliation[1] opens the door to the world of 2D materials. The linear dispersion and massless Dirac fermions yield unique electronic and optical properties of graphene[2-11]. Single layer graphene can absorb 2.3 % of the incident light from mid-infrared to the visible range[5, 6], which is originated from the universal optical conductivity $\sigma_0 = e^2/4\hbar$. As a 2D semimetal, the light absorption of graphene in far-infrared and terahertz range is dominated by free carrier response[11-13]. The strong and tunable light absorption of graphene provides platforms for a variety of optoelectronic and photonic applications such as photodetectors[14-16], optical modulators[17] and plasmonic devices[12, 13]. Subsequently, transition-metal dichalcogenides (TMDCs) ($MoS_2$, $MoSe_2$, $WS_2$, $WSe_2$ etc.) joined the family of 2D materials as key members. They possess indirect-to-direct band gap transitions as the thickness decreasing from bilayer to monolayer[18, 19]. The optical band gap of monolayer TMDCs is located at near-infrared to the visible region. Pronounced excitonic effects[20-23] and spin-valley coupling[24-27] have been demonstrated in TMDCs. Black phosphorus (BP) recently arouses much attention due to the extraordinary electronic properties[28, 29] and strongly layer-dependent band structures[30-33]. The strong exciton absorption peak covers a broad spectral range from mid-infrared to the visible.

InSe emerges as a competitive 2D semiconductor in recent years. In 2013, Mudd et al. exfoliated the bulk crystal of InSe into thin layers and revealed the largely layer-tunable band gap of few-layer InSe[34], which is among the pioneering studies regarding the atomically thin InSe. Since then, a large number of studies on the optical and electronic properties of few-layer InSe have appeared. In 2016, Bandurin et al. observed quantum Hall effect in high quality electronic devices based on few-layer

InSe[35], demonstrating its superior electron transport properties and further boosting the research interests. They also comprehensively studied the layer-dependent band structures by density functional theory (DFT) calculations and photoluminescent (PL) measurements, paving the way for future in-depth study. Few-layer InSe exhibits unusual optical properties compared to TMDCs and BP. The band gap transition is allowed for out-of-plane polarized light but forbidden or extremely weak for in-plane polarized light. Gisbert et al. performed *k*-space optical spectroscopy to determine the orientation of the luminescent excitons, showing that the light emission of few-layer InSe is mostly excited by the out-of-plane dipole[36]. Zultak et al. mapped the subband structure of InSe quantum well by resonant tunneling spectroscopy and photoluminescence excitation (PLE) measurements[37]. In addition to the study for the intrinsic physical properties, few-layer InSe has been fabricated into various optoelectronic devices and heterostructures in conjunction with other 2D materials. Zhao et al. designed a InSe/BP photodetector with polarized and fast photoresponse[38]. Ubrig et al. observed the robust interlayer excitons at the interface of InSe/TMDCs vdW heterostructures [39]. These studies show that few-layer InSe exhibits great potentials in electronic and optoelectronic applications owing to the unique electronic and optical properties.

In this perspective, we systematically review the studies on the optical properties of few-layer InSe, mostly from the perspective of experiments. There are four sections following this introduction. We start with the intrinsic optical and electronic properties of few-layer InSe. The layer-dependent band structures, light absorption and emission, Raman scattering and the electron transport properties will be introduced in section II. Then we survey several methods to tune the optical properties of few-layer InSe such as external strain, surface chemical doping and vdW interfacing in section III.

Furthermore, the applications of few-layer InSe in optoelectronics will be discussed in section IV. Finally, we present an outlook on the future study of the optical properties in section V.

**II. The intrinsic optical and electronic properties of few-layer InSe**

**A. Layer-dependent band structures**

The band structures of 2D materials can exhibit strong layer dependence due to the quantum confinement in the out-of-plane direction. For instance, there exists an indirect-to-direct bandgap transition for the TMDCs with thickness decreasing from bilayer to monolayer[18, 19]. Another well-known example is BP, possessing largely layer-tunable direct bandgap from the visible (the optical band gap is 1.73 eV for monolayer) to mid-infrared range (0.35 eV for bulk)[30-33]. The band structures of few-layer InSe can be largely modified by changing the layer number according to the DFT and/or tight binding (TB) calculations[40-47] (for a book see[48]) as well as PL experiments[34, 35, 49-54]. Fig. 1(a) shows the DFT band structure of monolayer InSe[43]. There are two interband transitions near the Γ point in Brillouin zone in which we are interested. One is the band gap transition between the valence band maximum (Se-pz orbital dominated) and the conduction band minimum (In-s orbital dominated), marked as transition-A. Another one is the interband transition between the deeper valence bands (degenerate Se-px/y orbitals dominated) and the conduction band (In-s orbital dominated), which is labeled as transition-B. For few-layer InSe, the valence/conduction band splits into subbands due to the interlayer interactions, resulting in a reduced band gap. Bandurin et al. systematically performed PL measurements on monolayer and few-layer InSe encapsulated by hexagonal boron nitride (hBN)[35]. Note that the PL peak is originated from the light emission of excitons, whose energy is the optical band gap. Optical band gap is smaller than the corresponding single-particle electronic band gap, where the difference corresponds to the binding energy[20]. As presented in Fig. 1(b), the optical band gap (transition-A) of InSe shifts from 1.25 eV for the bulk to about 1.9 eV for the bilayer. The reliable experimental determination of the optical band gap of monolayer InSe is still lacking. It is possibly at 2.1 eV according to another PL study[51]. The PL associated with the transition-B exhibits blueshift from 2.5 eV for 8-layer InSe to 2.9 eV for monolayer. The energy shift of the transition-A is larger than that of transition-B as the thickness changing. Because the valence band in transition-A is dominated by Se-pz orbital with orientation in the out-of-plane direction, this orbital is more sensitive to the interlayer coupling than the Se-px/y orbitals for transition-B.

The PL intensity of few-layer InSe decreases by more than an order of magnitude as the thickness decreasing from 10-layer to 2-layer[34, 35, 49-54], indicating a possible direct-to-indirect crossover. In light of DFT calculations, the highest valence band of monolayer InSe has a sombrero hat shape[40, 41], on the contrary to the parabolic shape for bulk InSe[55]. Hamer et al. measured the band structures (filled states) of monolayer and few-layer InSe by angle-resolved photoemission spectroscopy with submicrometer spatial resolution (micro-ARPES)[56]. As shown in Fig. 1(c), the highest valence band of monolayer is inversed at Γ point. The valence band maximum is located at $\Delta k = 0.3 \pm 0.1$ Å$^{-1}$ and $\Delta E = 50 \pm 20$ meV. As the thickness increases, more valence subbands emerge at the higher energy position than the original valence band (see Fig. 1(d)). For InSe thicker than 3 layers, the band inversion becomes vague since the energy of the band inversion is close to the thermal energy at room temperature ($k_B T$ = 26 meV for T = 300 K). The band inversion of 1-3 layers InSe is very small, so the band gap of them is actually "weak indirect" or "quasi direct". InSe eventually becomes a direct band gap semiconductor for 6 or more layers.

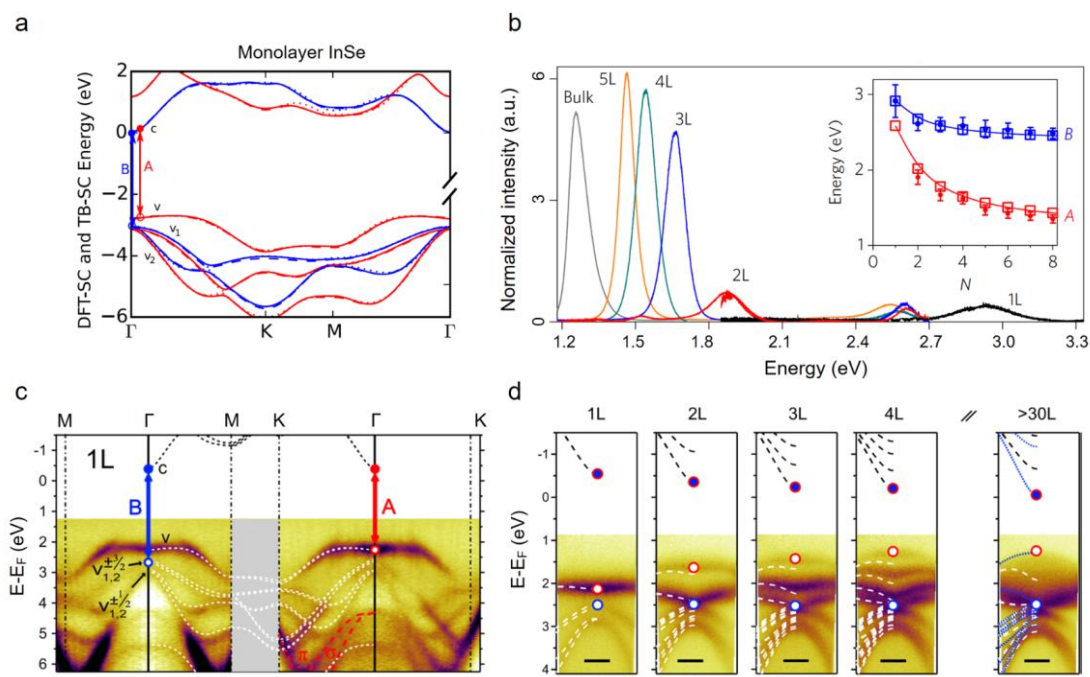

Fig. 1. (a) The DFT band structure of monolayer InSe[43]. Reproduced with permission from Phys. Rev. B 94, 245431 (2016). Copyright 2016 American Physical Society. (b) The layer-dependent PL spectra of InSe, the inset shows the PL energy for different layers[35]. Reproduced with permission from Nat. Nanotech. 12, 223 (2017). Copyright 2016, Springer Nature. (c) The band structure of monolayer InSe measured by micro-ARPES, (d) The evolution of valence bands as the thickness changing[56]. Reproduced with permission from ACS Nano 13, 2136 (2019). Copyright 2019 American Chemical Society.

Atomically thin 2D semiconductors with the presence of multiple subbands can be treated as natural quantum wells. For example, the optical transitions between different valence and conduction subbands are observed in few-layer BP[30-33]. Zultak et al. performed resonant tunneling spectroscopy to reveal the subband structures of few-layer InSe[37]. Few-layer InSe is sandwiched between two hBN flakes and connected by graphene electrodes. As shown in Fig. 2(a), the chemical potentials $\mu$ of the graphene electrodes are tuned by a bias voltage, which changes the band alignment of the InSe/graphene heterostructures. Once $\mu$ crosses the edge of a conduction subband, the tunneling current derivative ($dI/dV_b$) exhibits a sharp increase. As displayed in Fig. 2(b), there are 4 steps for 4-layer InSe with the energy difference of 0.5 eV between each conduction subband. The number of steps is consistent with the quantum number (n for n-layer quantum well). In addition, the optical transitions between different valence subbands ($c_n$) and the lowest conduction subband ($v_0$) are further confirmed by PLE as illustrated in Fig. 2(c). The splitting energy (5-layer InSe) between each valence subband is approximate 0.4 eV, which is consistent with the micro-ARPES results.

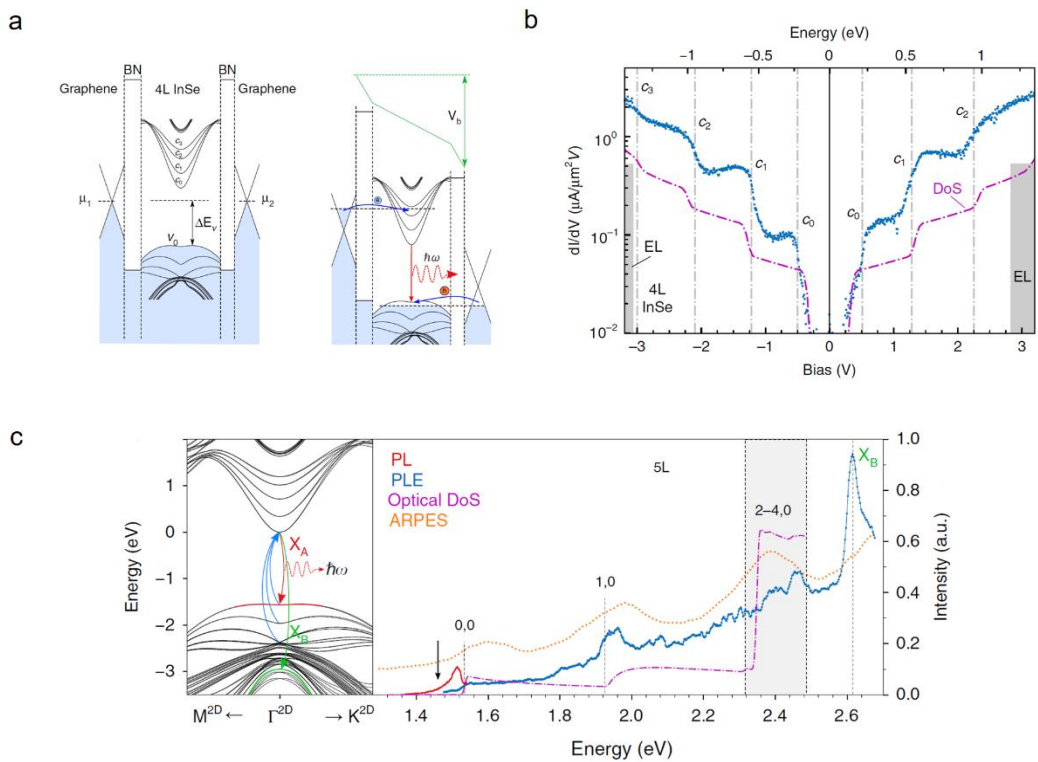

Fig. 2 (a) The band alignment of a 4-layer InSe device in unbiased and biased configurations. (b) The tunneling current derivative ($dI/dV_b$) under different bias voltages. (c) The PLE spectrum of a 5-layer InSe (blue line)[37]. Reproduced with permission from Nat. Commun. 11, 125 (2020).

### B. Light absorption, emission and excitonic effect

Resembling other 2D semiconductors like TMDCs and BP, the light absorption and emission of few-layer InSe exhibit the same layer dependence as the band structures, which has been confirmed by PL measurements[34, 35, 49-54]. However, the light-matter interaction of few-layer InSe is fundamentally different from those of TMDCs and BP. Because the band gap transition of few-layer InSe is inactive or extremely weak for in-plane polarized light. According to the Fermi's golden rule, the absorption coefficient of light for the transition from the valence band to the conduction band is given by[57]:

$$\alpha_{cv}(\omega) = \frac{e^2}{\pi n c m_e^2 \omega} \int \left|\vec{P_{cv}}(\vec{k})\right|^2 \delta(\hbar\omega - E_g) d\vec{k} \qquad (1)$$

Where $n$ is the refractive index of the material, $m_e$ is electron mass and $E_g$ is the energy of the interband transition. The dipole matrix element is expressed as $\vec{P_{cv}}(\vec{k}) = \langle \vec{k}, c | \vec{e} \cdot \vec{P} | \vec{k}, v \rangle$, $\vec{e}$ is the unit polarization vector, $\vec{P}$ is the interband momentum operator. $\delta(\hbar\omega - E_g)$ is proportional to the joint density of states. According to equation (1), the light absorption of solids is determined by dipole matrix element and the joint density of states. The in-plane dipole matrix elements vanish for monolayer InSe. Monolayer InSe has $z \rightarrow -z$ mirror symmetry (the crystal structure of γ-phase InSe is displayed in the next section), so the wave function of valence/conduction band is either even or odd with respect to the mirror plane. Without considering the spin-orbit interaction, the wave functions for the highest valence band (Se-pz orbitals dominated), the deeper valence bands (Se-px/y orbitals dominated) and the lowest conduction band (In-s orbitals dominated) are even, odd and odd respectively with respect to the $z \rightarrow -z$ mirror symmetry[41, 43]. The absorption of in-plane (out-of-plane) polarized light requires the wave functions of valence band and conduction band have the same (opposite) $z \rightarrow -z$ symmetry. Therefore, for monolayer InSe, the band gap transition (transition-A) is forbidden for in-plane polarized light and allowed for out-of-plane polarized light, while for transition-B the situation is the opposite. The selection rules persist for few-layer and bulk InSe although the $z \rightarrow -z$ symmetry is broken with the change of stacking order[55]. Magorrian et al. calculated the dipole momentum matrix of monolayer and few-layer InSe by $k \cdot p$ perturbation theory[43]. For the band gap transition of few-layer InSe, the in-plane dipole matrix elements are zero at $\Gamma$ point in Brillouin zone and increase linearly at small momentum as displayed in Fig. 3(a). The out-of-plane dipole matrix element for the band gap transition is always nonzero for monolayer and few-layer InSe. Consequently for transition-A, monolayer and few-layer InSe can only efficiently absorb light with out-of-plane polarization. For transition-B, the in-plane dipole matrix element is nonzero, hence it can be excited by in-plane polarized light. We measured the differential reflectance spectra $(R-R_s)/R$ of monolayer and few-layer InSe[58], which is proportional to the absorption coefficient $\alpha(\omega)$ of light for the materials as follows[59, 60]:

$$\frac{R-R_s}{R} = \frac{4n}{n_s^2 - 1} \alpha(\omega) \quad (2)$$

Where $R$ and $R_s$ correspond to the reflected light from the material and the substrate, n and $n_s$ are the refractive indices of the material and the substrate respectively. As illustrated in Fig. 3(b), the absorption of in-plane polarized light near the band gap is extremely weak. For example, the optical band gap of bilayer InSe is 1.9 eV according to the PL measurements[35], but the differential reflectance spectrum is featureless at this spectral range. In contrast, the layer-dependent exciton peak associated with transition-B can be clearly observed.

Although the band gap transition of few-layer InSe is forbidden for in-plane polarized light, the PL emission mainly contributed from the out-of-plane polarized light can be measured[34, 35, 49-54]. Gisbert et al. detected stronger PL emission of InSe flakes on the ramp of $SiO_2$ nanoparticle (5 times) than that on flat region[51], as illustrated in Fig. 3(c). Similar phenomena were observed for bended InSe[61] and the ridges of wrinkled InSe[62]. The propagation direction of the light excited by the out-of-plane dipole is mainly parallel to the plane direction, therefore stronger PL signal can be collected along this direction. It should be noted that, however, the strain induced modification of band structures and electronic properties may exist in these inhomogeneous samples[54, 63], hence the enhancement of PL signal may not be purely originated from the out-of-plane dipole. In the following, Gisbert et al. performed *k*-space optical spectroscopy to unambiguously determine the orientation of the PL emission[36]. They found that 95% of the PL emission for few-layer InSe is originated from the out-of-plane dipole as demonstrated in Fig. 3(d). The weak emission from the in-plane dipole is due to the mixing between the Se-pz orbitals dominated valence band and the Se-px/y orbitals dominated valence band when spin-orbit interactions are induced, which has been concurrently confirmed by their simulations. In another theoretical paper[45], the absorption coefficient of in-plane polarized light is estimated to be 1.5 % for monolayer and ~ 0.3 % for few-layer InSe when spin-orbit couplings are considered. The dominant light emission originated from the out-of-plane dipole can be confirmed by the angle-dependent PL spectroscopy for the lamella-shaped InSe samples[56]. As shown in Fig. 3(e), the PL intensity is maximal when the detected polarization is perpendicular to the basal plane of InSe layers.

Strong excitonic effect was observed for bulk InSe at low temperature[64-68]. Exciton is an electron-hole pair held together by Coulomb interaction. For bulk InSe at liquid-helium temperature, the ground state (1s) and the first excited state (2s) of exciton near the band edge can be resolved by transmission measurements[64]. The 1s state of exciton is at about 1.353 eV and the binding energy is 14.5 meV. Recently, Shubina et al. performed temperature-dependent PL measurements on bulk InSe[68]. A series of PL peaks were observed by lowering temperature to 10 K, as shown in Fig. 3(f). There are three exciton-related PL peaks, which are free exciton at 1.338 eV (X), biexciton at 1.335 eV (M), and P-band of the exciton-exciton (X-X) scattering at 1.320 eV (P). The nature of the biexciton and X-X scattering are confirmed by quadratic power dependence. The PL peaks below 1.3 eV correspond to the emission from a series of defect states[69, 70]. For few-layer InSe, the defect-related PL peaks become much stronger (up to 60 times) than the PL of free exciton at low temperature[61, 62], because atomically thin 2D materials are more vulnerable to the external environment. The exciton binding energy of bulk InSe is determined by the energy difference between the free exciton and the P-band X-X scattering, which is about 20 meV. For atomically thin 2D materials, the exciton binding energy generally increases as the thickness decreases due to the confinement and the weakening of screening[20]. For example, the exciton binding energy of BP increases from 106 meV for 6-layer to 213 meV for bilayer[32]. Although InSe has been demonstrated as a strongly layer-dependent semiconductor, the direct observation of the ground state and excited state of exciton is challenging due to the transition selection rules as discussed above. Instead, Zultak et al. determined the binding energy of few-layer InSe by PLE spectroscopy[37]. The incident light is both in-plane and out-of-plane polarized for lamella-shaped InSe samples. According to the energy difference between the exciton peak and the onset of continuum absorption in PLE spectrum, the binding energy of 2-layer InSe is estimated to be about 50 meV as illustrated in Fig. 3(g).

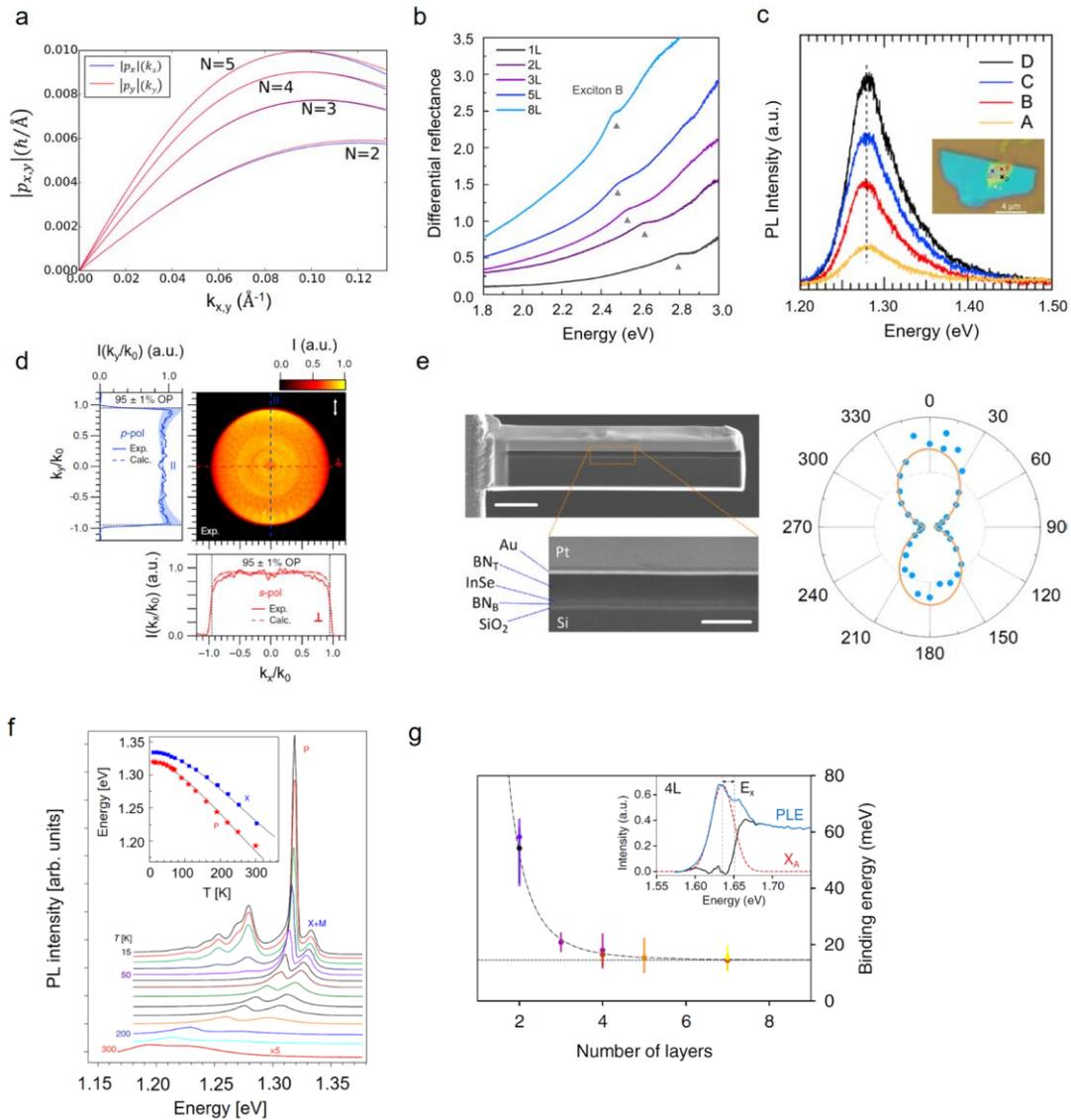

Fig. 3 (a) The in-plane dipole matrix elements for the band gap transition of few-layer InSe[43]. Reproduced with permission from Phys. Rev. B 94, 245431 (2016). Copyright 2016 American Physical Society. (b) The differential reflectance spectra of monolayer and few-layer InSe[58]. Reproduced with permission from Phys. Rev. B 99, 195414 (2019). Copyright 2019 American Physical Society. (c) The PL spectra of InSe on the ramp of $SiO_2$ nanoparticles[51]. Reproduced with permission from Nano Lett. 16, 3221 (2016). Copyright 2016 American Chemical Society. (d) The k-space emission pattern of an 8-nm-thick InSe[36]. Nat. Commun.10, 3913 (2019). Copyright 2019

Springer Nature. (e) The scanning electron micrographs and the angle-dependent PL of the lamella-shaped InSe. 90° corresponds to the basal plane of the InSe layer[56]. Reproduced with permission from ACS Nano 13, 2136 (2019). Copyright 2019 American Chemical Society. (f) The temperature-dependent PL spectra of bulk InSe[68]. Reproduced with permission from Nat. Commun. 10, 3479 (2019). Copyright 2019 Springer Nature. (g) The layer-dependent binding energy of InSe exciton determined by PLE spectroscopy[37]. Reproduced with permission from Nat. Commun. 11, 125 (2020). Copyright 2020 Springer Nature.

### C. Crystal structure and Raman scattering

There are at least three crystal phases for InSe, which are γ, β and ε phase[71, 72]. The single layer of them have the same structure but the stacking order between each layer is different. The widely studied γ-phase InSe has a rhombohedral structure as illustrated in Fig. 4(a). In each layer, it has honeycomb lattice with four atomic planes arranged in the sequence of Se-In-In-Se, and the atoms within the layer are connected by covalent bonds. The thickness of monolayer InSe is about 0.8 nm[35]. The single layers are held together by vdW forces, with the stacking order that the selenium atoms in one layer are on top of the indium atoms of the layer beneath (ABC staking order). Few-layer InSe (γ-phase) belongs to point group $C_{3v}$, while monolayer InSe belongs to point group $D_{3h}$ with additional $z \rightarrow -z$ mirror symmetry. The broken inversion symmetry gives rise to high nonlinearities as in the optical second harmonic generation of monolayer and few-layer InSe[73-75]. The normal modes of vibrations can be decomposed as $4A_1$ ($\Gamma_1$) + $4E$ ($\Gamma_3$). One $A_1$ and one E modes are acoustic, while the others are optical modes which are both Raman and infrared active[76]. The vibrational structures of the optical modes are displayed in Fig. 4(b). Among them, the $A_1$ ($\Gamma_1^1$) and E ($\Gamma_3^1$) modes split into longitudinal optical (LO) modes and transverse optical (TO) modes. Raman spectroscopy is performed to study the vibrational properties of few-layer InSe[49, 53, 58, 77, 78]. The frequency of E ($\Gamma_2^3$) mode is about 41 cm$^{-1}$, which is beyond the lower limit of common Raman spectroscopy[76, 79]. The other phonon modes are A$_1$ ($\Gamma_1^2$) at 115 cm$^{-1}$, E ($\Gamma_3^3$)/E ($\Gamma_1^3$)-TO at 178 cm$^{-1}$ (these two modes have close or degenerate energy), A$_1$ ($\Gamma_1^1$)-TO at 196 cm$^{-1}$, A$_1$ ($\Gamma_1^1$)-LO at 199 cm$^{-1}$ and A$_1$ ($\Gamma_3^1$) at 227 cm$^{-1}$ for 30-layer InSe[58]. Sánchez-Royo et al. demonstrated that the phonon frequency of InSe exhibits thickness dependence[49]. The frequency of A$_1$ ($\Gamma_1^2$) mode slightly decreases (by 0.5 cm$^{-1}$) and those of E ($\Gamma_3^3$) and A$_1$ ($\Gamma_1^1$)-LO modes prominently increase (by 0.7-1.5 cm$^{-1}$) as thickness decreasing from 12-layer to 6-layer. This shows that the In-Se bonds become stronger for thinner layers, whereas it is the opposite for In-In bonds. The thickness dependent phonon frequency can be utilized in determining the layer number.

The A$_1$ ($\Gamma_1^1$)-LO and E ($\Gamma_1^3$)-LO phonon modes of InSe are forbidden according to selection rules or relatively weak in the backscattering geometry[80]. But the intensity of Raman scattering can be largely enhanced when the energy of the incident photon is close to that of transition-B. For bulk and thick InSe flakes (over 10 layers), the energy of transition-B is about 2.45 eV at room temperature[58]. Therefore, the Raman scattering is under near-resonant conditions and the A$_1$ ($\Gamma_1^1$)-LO and E ($\Gamma_1^3$)-LO modes are observable with the excitation of 514.5 nm laser (2.41 eV). In contrast, these two phonon modes are not observed when Raman scattering is excited by 532 nm laser (2.33 eV) or 633 nm laser (1.96 eV)[58, 77]. The energy of transition-B can be tuned by temperature, thickness and strain, then the resonance condition is changed accordingly. For example, at liquid-nitrogen temperature, bulk InSe exhibits stronger resonance effect under the excitation of 488 nm laser (2.54 eV) than excited by 514.5 nm laser (2.41 eV)[81-83]. It manifests the temperature dependence of transition-B energy. For few-layer InSe, the energy of transition-B exhibits strong layer dependence. It shifts from about 2.45 eV for 10 or more layers InSe to 2.9 eV for monolayer InSe (at room temperature)[35, 58]. Therefore, as shown in Fig. 4(c), the Raman intensity of $A_1$ ($\Gamma_1^1$)-LO mode gradually decreases as the thickness is reduced when the excitation photon energy is fixed at 2.41 eV (514.5 nm)[49, 77]. In addition, strain can effectively tune the energy of transition-B, and thus change the resonance effect of Raman scattering[58]. As shown in Fig. 4(d), the $A_1$ ($\Gamma_1^1$)-LO and E ($\Gamma_1^3$)-LO modes are largely enhanced by an order of magnitude when 1.15 % uniaxial tensile strain is applied to 10-15 layers InSe. PL measurements indicate that the energy of transition-B redshifts by 50 meV under such magnitude of strain, making the energy of transition-B perfectly match the energy of incident photon (fixed at 2.41 eV). Overall, resonant Raman spectroscopy is a versatile tool to study the vibration and electronic properties as well as the electron-phonon interactions of few-layer InSe.

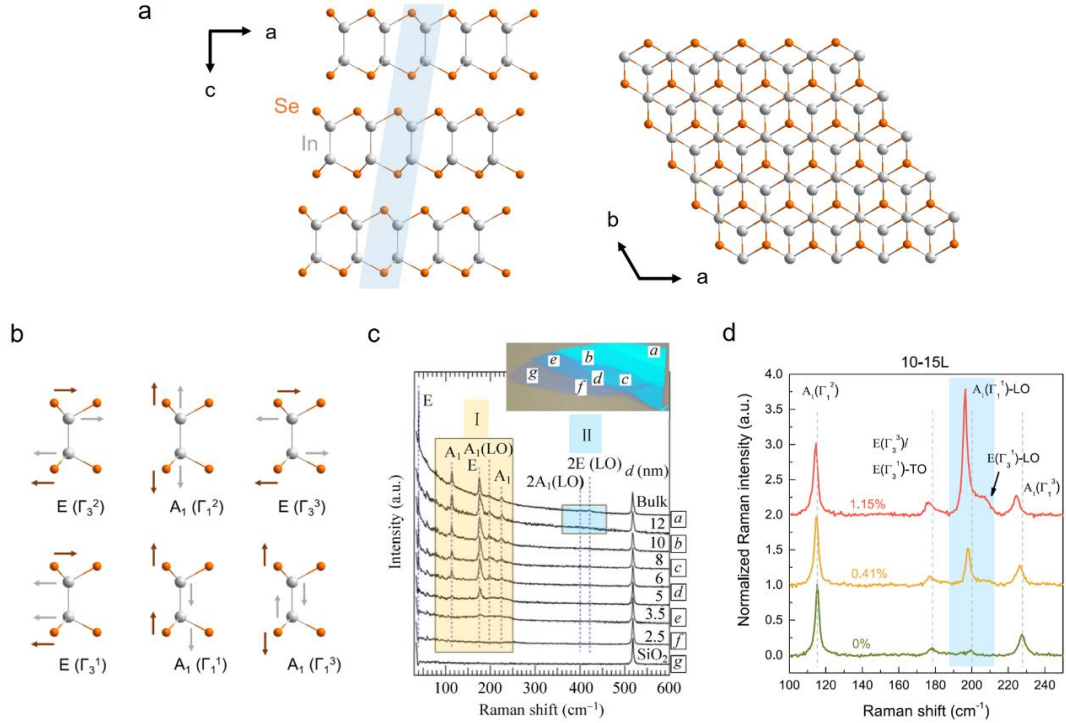

Fig. 4 (a) The crystal structure of γ-phase InSe viewed from b-axis direction (left) and c-axis direction (right). (b) The vibrations of optical phonon modes. (c) The layer-dependent Raman spectra of InSe[49]. Reproduced with permission from Nano Res. 7, 1556 (2014). Copyright 2014 Springer Nature. (d) The strain-dependent Raman spectra of few-layer InSe[58]. Reproduced with permission from Phys. Rev. B 99, 195414 (2019). Copyright 2019 American Physical Society.

### D. High carrier mobility

Few-layer InSe exhibits high carrier mobility as a 2D semiconductor. A large number of high performance field effect transistors (FETs)[35, 84-96] and photodetectors[38, 77, 97-117] were fabricated based on few-layer InSe. Bandurin et al. demonstrated the superior electron transport properties of few-layer InSe encapsulated by hBN[35]. By applying top and bottom gate voltages, the carrier density can be tuned over a wide range from $10^{12}$ to $10^{13}$ cm$^{-2}$. As shown in Fig. 5(a), the Hall mobility of a 6-layer InSe is about 1000 cm$^2$/Vs at room temperature and increases as temperature decreasing.

The highest mobility is 12700 cm$^2$/Vs at liquid-helium temperature with carrier density n ≈ 8 × 10$^{12}$ cm$^{-2}$. Quantum Hall effect is observed based on InSe electronic devices as displayed in Fig. 5(b). In addition, high carrier mobility can be achieved by fabricating InSe electronic devices on PMMA/Al$_2$O$_3$ substrates as shown in Fig. 5(c)[84, 85]. Because the scattering with interfacial Coulomb impurities is suppressed by the dielectric screening of PMMA. Feng et al. demonstrated that the FET mobility of few-layer InSe on PMMA/Al$_2$O$_3$ substrates is prominently enhanced (1055 cm$^2$/Vs at room temperature) compared to that of InSe on Al$_2$O$_3$ substrates (64 cm$^2$/Vs) (see Fig. 5d)[84]. The current on/off ratio is up to 1×10$^8$. In addition, the carrier mobility of few-layer InSe can be enhanced by dry oxidation[86, 93], surface charge doping with indium layers[88, 92] and Sn-doping[114] etc. In summary, the high carrier mobility and high current on-off ratio make few-layer InSe promising for electronic and optoelectronic applications.

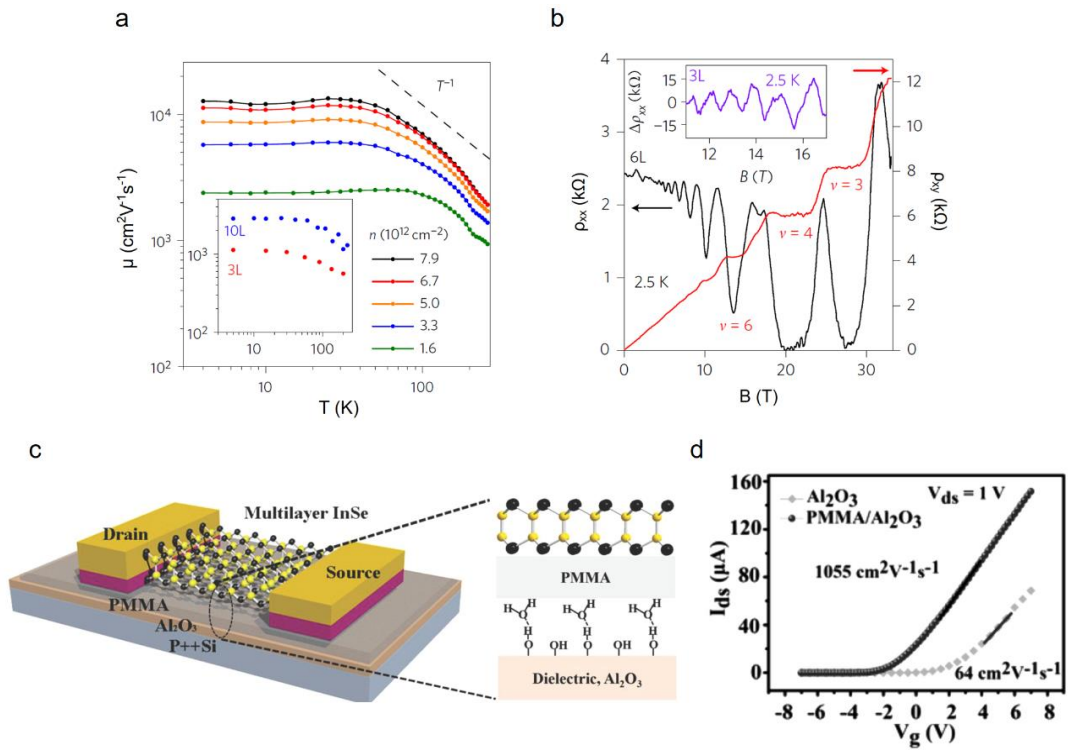

Fig. 5 (a) Temperature dependent Hall mobility for a 6-layer device. (b) Quantum Hall effect for the 6-layer device[35]. Reproduced with permission from Nat. Nanotech. 12, 223 (2017). Copyright 2016 Springer Nature. (c) The schematic illustration of an InSe FET on PMMA/Al$_2$O$_3$ substrates. (d) The transport characteristics of few-layer InSe FETs on PMMA/Al$_2$O$_3$ and Al$_2$O$_3$ substrates[84]. Reproduced with permission from Adv. Mater. 26, 6587 (2014). Copyright 2014 WILEY-VCH Verlag GmbH & Co. KGaA, Weinheim.

### III. Tuning the optical properties of few-layer InSe

#### A. External strain

2D materials exhibit great stretchability compared to their bulk counterpart. Therefore, the vibrational, optical and electronic properties of 2D materials can be effectively tuned by strain[118]. The strain effect on 2D materials such as graphene[119, 120], TMDCs[121-123] and BP[31, 124, 125] have been extensively studied. Zhao et al. demonstrated that few-layer InSe has superior flexibility, because its Young's modulus (23.1 ± 5.2 GPa)[126] is much lower than those of other 2D materials like graphene (~1000 Gpa)[127] and TMDCs (~400 Gpa)[128]. DFT calculations predict an indirect-to-direct band gap transition for monolayer InSe under 6% uniaxial compress strain[129, 130]. We observed the largely tunable band structures of few-layer InSe by strain according to PL measurements[54]. As illustrated in Fig. 6(a), uniaxial tensile strain is applied continually and reversely by bending the flexible substrates such as polypropylene (PP) or poly (ethyleneterephthalate) (PET). Fig. 6(b) shows the PL spectra of a strained 5-layer InSe. The PL peak associated with transition-A shifts from 1.49 eV to 1.38 eV when uniaxial tensile strain is applied from zero to 1.15%. The shift rates is about 100 meV/%, which is larger than that of monolayer MoS$_2$ (45-70 meV/%)[121, 122] and close to that of BP (100-200 meV/%)[125]. Li et al. reported a higher shift rate for few-layer InSe (150

meV/%)[63]. The redshift of band gap induced by tensile strain can be interpreted by DFT calculations. The in-plane tensile strain can weaken the In-Se bond strength and thus raise the bonding valence band (Se-pz oribital dominated) and lower the antibonding conduction band (In-s orbital dominated). It is usually difficult to tune the properties of bulk materials by stretching method. However, by taking advantage of superior flexibility, the band structures of bulk-like InSe (over 50 layers) can be effectively engineered by strain (see Fig. 6(c)). The shift rates are 118 meV/% for transition-A and 43 meV/% for transition-B. The shift rate of transition-B is smaller than that of transition-A because Se-px/y orbital bands remain almost unchanged under strain. In addition to the modification of band structures, the vibration properties of few-layer InSe can be tuned by strain. Tensile strain weakens the restoring force in vibrations and thus reduces phonon frequency. Phonon softening is observed for few-layer InSe. The shift rate is at the range of 0.8-2.5 cm$^{-1}$/% for different phonon modes[58]. Since strain can efficiently tune the optical and electronic properties of few-layer InSe, the performance of flexible electronic and optoelectronic devices based on few-layer InSe can exhibit large tunability by strain[63, 98, 116, 131]. For example, Dai et al. demonstrated the piezo-phototronic effect in flexible photodetectors based on 30 nm-InSe as presented in Fig. 6(d)[116]. The responsivity and the response speed of the photodetector are enhanced by nearly an order of magnitude when 0.62 % uniaxial tensile strain is applied.

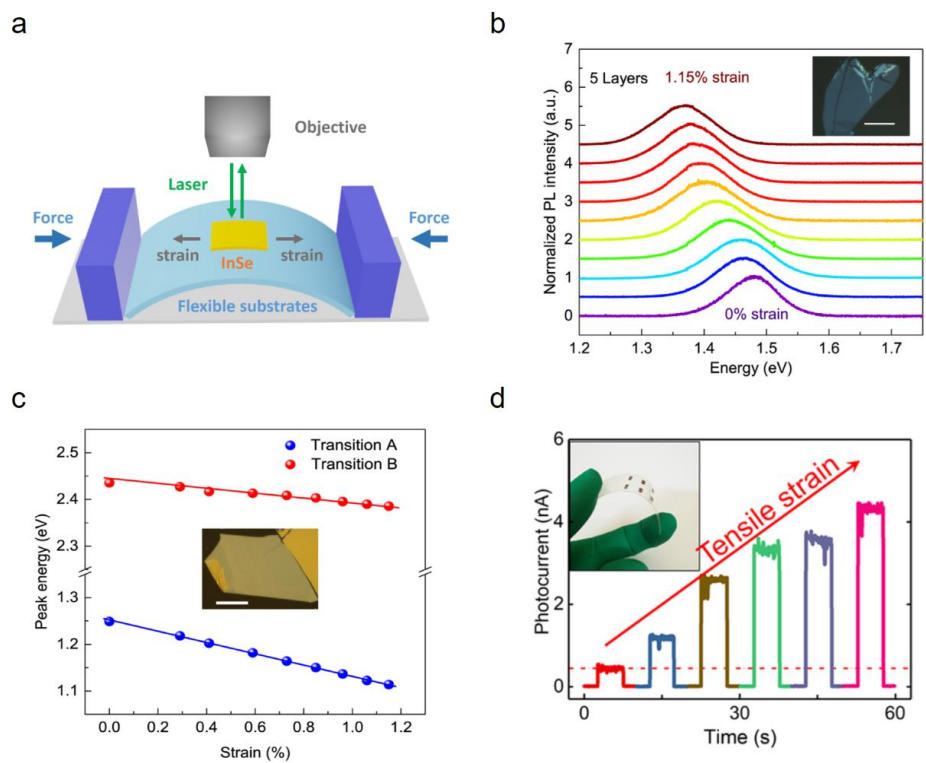

Fig. 6 (a) Schematic illustration of the two-point bending method to apply uniaxial tensile strain. (b) The evolution of the PL spectra of a 5-layer InSe under uniaxial tensile strain. (c) The energy evolution of PL peaks associated with transition-A and transition-B, the inset shows the photograph of a bulk-like InSe (over 50 layers)[54]. Reproduced with permission from ACS Appl. Mater. Interfaces 10, 3994 (2018). Copyright 2018 American Chemical Society. (d) The photocurrent of a flexible InSe photodetector, the inset shows the image of the flexible device[116]. Reproduced with permission from ACS Nano 13, 7291 (2019). Copyright 2019 American Chemical Society.

### B. Surface chemical doping

The band gap of 2D materials can be manipulated by surface chemical doping. The formation of surface dipole layer can induce electrical field and thus lead to the surface stark effect[132, 133]. Zhang et al. performed ARPES to directly observe the band gap renormalization of InSe by depositing Na atoms[134]. As shown in Fig. 7(a), there is no signal of valence band for pristine InSe. As the deposition time increases, the valence band is clearly observed and shifts downwards. Then, the band gap of InSe can be directly determined by the distance between valence band maximum and conduction band minimum. As illustrated in Fig. 7(b), the band gap of InSe is reduced by 120 meV as the carrier concentration increases to $8.1\times10^{12}$ cm$^{-2}$ after depositing for 17 min. X-ray photoemission spectra (XPS) confirms that the deposition process is dominated by the surface absorption of Na atoms rather than intercalation into the layers. The effect of surface doping is similar to that of applying gate voltages in FETs. The moderate carrier concentration possibly can be realized by electrical gating.

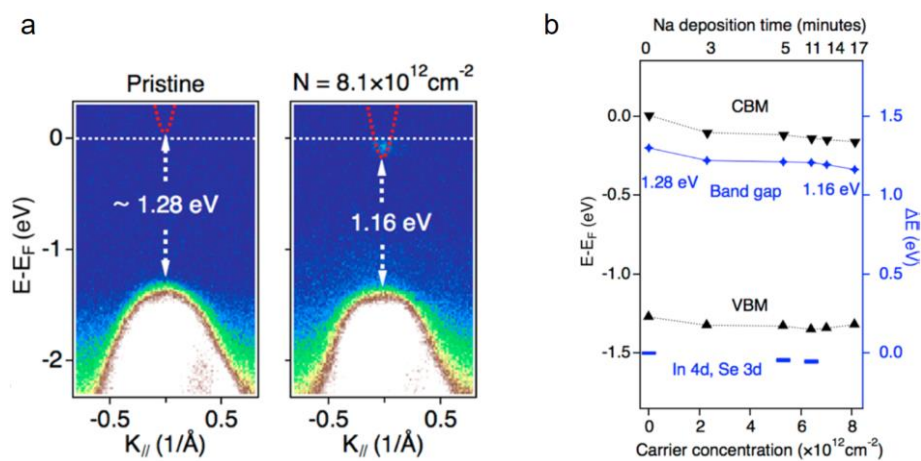

Fig. 7 (a) The band structure of InSe without and with surface Na doping measured by ARPES. (b) The band gap evolution as a function of carrier concentrations[134]. Reproduced with permission from ACS Nano 13, 13486 (2019). Copyright 2019 American Chemical Society.

### C. Interlayer excitons at vdW interfaces

Van der Waals heterostructures can be prepared by stacking various 2D materials. Since the materials are weakly coupled by vdW interactions, one can freely choose materials to achieve desired physical properties[135, 136]. Few-layer InSe has been fabricated into vdW heterostructures with various other 2D materials like GaSe[97, 104], GaTe[108, 117], graphene[111, 137], BP[38, 95, 96, 112, 113] etc. Most of these studies mainly focus on the electronic and optoelectronic devices of InSe heterostructures, while the basic optical properties of them are rarely mentioned. We will specifically introduce them in the next section. Herein, we highlight one recent study regarding the interlayer exciton at the InSe/TMDCs interface[39]. For heterostructures with type-II band alignment, the formation of interlayer exciton will dramatically modify the optical properties compared to the pristine materials. Although interlayer excitons can in principle exist in type-II interface, the efficient light emission from the recombination of them requires the transition to be direct in the momentum ($k$) space[138]. Specifically, the conduction band minimum of one layer should be at the same position in $k$ space as the valence band maximum in the other layer. Ubrig et al demonstrated that the vdW interface of InSe/TMDCs can support robust interlayer exciton[39]. As shown in Fig. 8(a), the bilayer InSe/ $WS_2$ heterostructure exhibits type-II band alignment. The valence band maximum is in $WS_2$ layer and the conduction band minimum is in InSe layer, both located at the $\Gamma$ point (k = 0) in Brillouin zone. As displayed in Fig. 8(b), the PL peak of the interlayer exciton is at 1.55 eV (T = 5K), while the PL peak from the free exciton of bilayer InSe ($WS_2$) is at 1.9 eV (1.73 eV). Fig. 8(c) shows the PL of the interlayer excitons for multiple-layer InSe/2-layer $WS_2$ heterostructures. The PL peak redshifts from 1.55 eV for the 2-layer InSe/2-layer $WS_2$ heterostructure to 1.25 eV for the 7-layer InSe/2-layer $WS_2$ heterostructure. The band alignments of them are presented in Fig. 8(d). The energy of the direct interlayer exciton transitions can be further tuned by varying the thickness of $WS_2$ or replacing $WS_2$ by other TMDCs like $WSe_2$, $MoS_2$ and $MoSe_2$. These studies demonstrate that InSe/TMDCs heterosturctures exhibit great potentials for broadband optoelectronic devices.

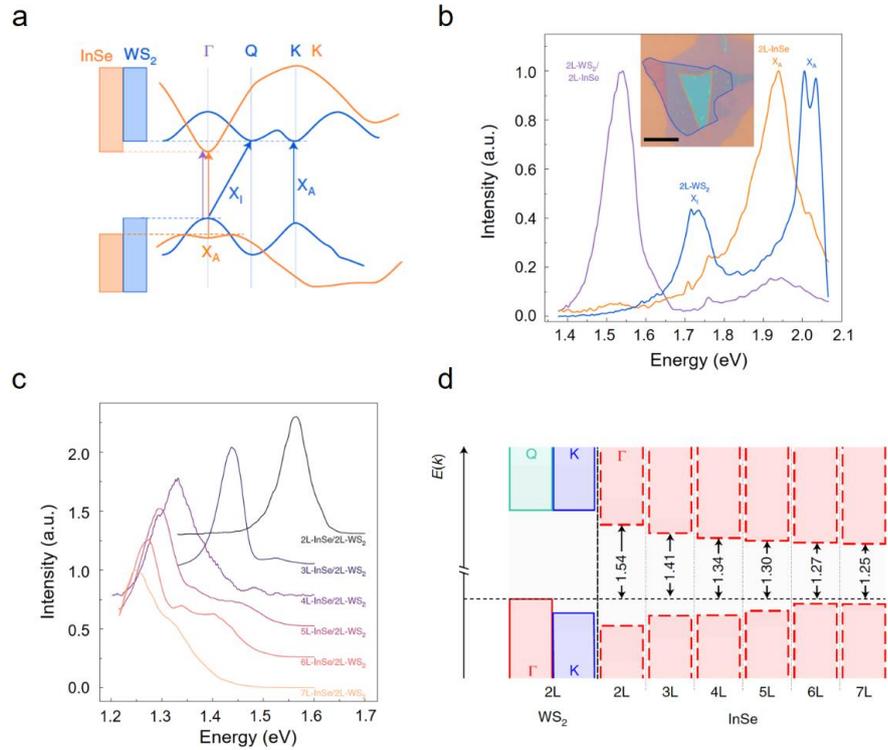

Fig. 8 (a) The band structure and the band alignment of a bilayer InSe/WS$_2$ heterostructure. (b) The PL of the interlayer exciton for the 2-layer InSe/2-layer WS$_2$ heterostructure and the intralayer excitons of prinstine 2-layer InSe and 2-layer WS$_2$. (c) The PL spectra and (d) the band alignments of multiple-layer InSe/2-layer WS$_2$ heterostructures[39]. Reproduced with permission from Nat. Mater. 19, 299 (2020). Copyright 2020 Springer Nature.

## IV. The applications based on few-layer InSe

Few-layer InSe can be utilized in a variety of electronic and optoelectronic devices, such as FETs[35, 84-96], photodetectors[38, 77, 97-117], photosensors[139, 140], solar cells[61] and thermoelectric devices[141]. In this section, we will focus on the photodetectors based on few-layer InSe. In part A, the InSe photodetectors will be introduced, while in part B the photodetectors based on the heterostructure of InSe with other 2D materials will be surveyed.

### A. Photodetectors

Benefiting from the layer-dependent band gap and the excellent electronic properties, few-layer InSe is suitable for high performance broadband photodetectors[38, 77, 97-117] (for other review see[142, 143]). Tamalampudi et al. fabricated an photodetector based on the mechanically exfoliated few-layer InSe[98] as illustrated in Fig. 9(a). The InSe photodetector exhibits broad spectral photoresponse from the visible to near-infrared region (450-785 nm). As shown in Fig. 9(b), the photoresponsivity and external quantum efficiency (ECE) increase as the wavelength of incident light decreasing. Under the illumination of 633 nm laser, the photoresponsivity is about 6.9 AW$^{-1}$ (P = 2.1 mWcm$^{-2}$, $V_{ds}$ = 10 V and $V_g$ = 0 V), which is much larger than those of graphene photodetectors (~ 0.5 mAW$^{-1}$)[14] and MoS$_2$ photodetectors (~ 7.5 mAW$^{-1}$)[144]. The ECE and the detectivity are estimated to be ~1367% and 1.07 × 10$^{11}$ Jones respectively under this condition. The photoresponse can be efficiently enhanced by electrical gating, because the Fermi level of InSe can be tuned by gating and thus reduce the barrier between the Fermi level of Au electrodes and the conduction band of InSe. As shown in Fig. 9(c), the photoresponsivity increases from 6.9 AW$^{-1}$ to 157 AW$^{-1}$ as the $V_g$ is swept from zero to 70 V. The dynamic response of the InSe photodetector is determined by the on-off switching of light illumination (λ = 633 nm, P = 350 mWcm$^{-2}$ and $V_{ds}$ = 2.0 V). The rise and fall time is about 40-60 ms as demonstrated in Fig. 9(d), which is comparable to those of GaSe photodetectors(~ 20 ms)[145] and MoS$_2$ photodetectors[144] (~ 50 ms).

The performance of InSe photodetector can be improved by using graphene instead of Au as electrodes[101-103]. The work function of Au ($W_{Au}$ = 5.2 eV) is larger than that of n-type InSe ($W_{InSe}$ = 4.8 eV), leading to the Schottky contact at the InSe/Au interface[102]. In contrast, the work function of graphene is smaller ($W_G \approx$ 4.5 eV), and it can be tuned by electrical gating. Therefore, good Ohmic contacts can be realized at the InSe/graphene interface. In addition, graphene and InSe can be vertically stacked to form a photodetector, because graphene is nearly transparent and plays little role on the absorption of light at the near-infrared and visible range. In this vertical structure, the distance between source and drain is largely reduced, hence the transit time of carrier is decreased and the photoresponse becomes faster. Mudd et al. fabricated two kinds of vertical graphene/InSe photodetectors[102] as illustrated in Fig. 9(e). The maximum photoresponsivity of them is close to $10^5$ AW$^{-1}$ ($\lambda$ = 633 nm, P = $10^{-14}$ W over the area of InSe flake ~ 16 μm$^2$, $V_{ds}$ = 2 V, $V_g$ = 0V) as illustrated in Fig. 9(f), which is dramatically enhanced compared to InSe detectors with Au electrodes.

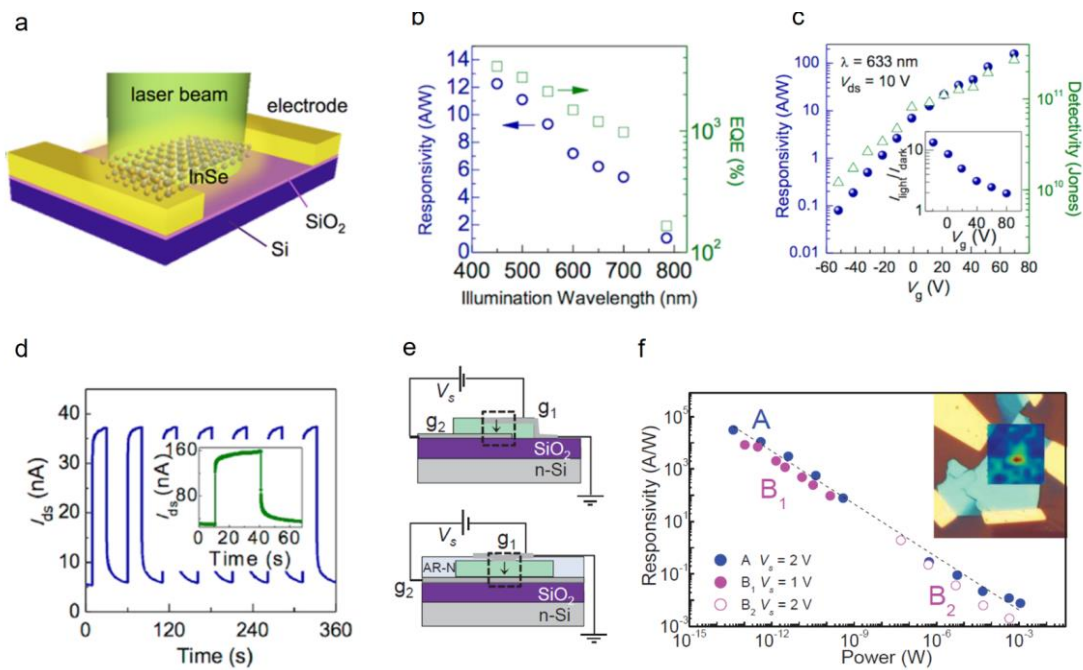

Fig. 9 (a) The schematic illustration of an InSe photodetector. (b) The wavelength-dependent photoresponsivity, (c) the photoresponsivity under different gate voltage and (d) the dynamic photoresponse of the InSe photodetector[98]. Reproduced with permission from Nano Lett. 14, 2800 (2014). Copyright 2014 American Chemical Society. (e) The schematic illustration of vertical InSe/graphene photodetectors. (f) The power-dependent photoresponsivity of the vertical InSe/graphene photodetectors[102]. Reproduced with permission from Adv. Mater. 27, 3760 (2015). Copyright 2015 the Authors. Published by WILEY-VCH Verlag GmbH & Co. KGaA, Weinheim.

Plasmonic devices can enhance the light absorption and thus improve the performance of InSe photodetectors[107, 115]. Dai et al. fabricated a self-powered InSe photodetector coupled with Au surface plasmon[107]. As shown in Fig. 10(a), asymmetric Schottky junction was fabricated by using Au as the electrode in one side and In in the other. Because of the different work function for Au and In, the photon-generated carriers can be separated by the built-in electric field without a bias voltage. Triangle Au plasmonic nanoparticles (NPs) array was patterned onto the surface of InSe flake. The side length of the triangle Au NPs is about 70-250 nm and the thickness is 25 nm. There are two absorption peaks according to the extinction spectrum as shown in Fig. 10(b), which are attributed to the surface plasmon of the triangle Au NPs originated from the dipole (1080 nm) and the quadrupole (664 nm). The absorption peaks of Au surface plasmon give rise to the enhancement of photoresponsivity. As displayed in Fig. 10(c), the photoresponsivity exhibits a peak in the spectral range of 580-800 nm. The photoresponsivity of the Au plasmon/InSe photodetector is enhanced by 1200% in this range compared to the InSe detector without coupling to the surface plasmon. In addition, the detectivity is enhanced and the rise/fall time is reduced (from ~50 ms to ~25 ms) in the meantime.

In addition to using graphene as electrodes and coupling to plasmonic devices, the photoresponse of InSe detector can be modified by external strain[63, 98, 116, 131], surface oxidation[106], chemical doping[114], avalanche effect[100, 115] and liquid-phase exfoliation[109, 110] etc. Among them, the lateral size of the exfoliated InSe is limited to the order of micrometers, which is difficult for practical applications. Large-scale InSe can be synthesized by chemical vapor transport/deposition[94, 146] and pulsed laser deposition (PLD)[105] methods. Yang et al. grew wafer-scale InSe (ε-phase) thin films by PLD[105]. The InSe thin film is centimeter-scale as shown in Fig. 10(d). It exhibits good uniformity and high crystallinity according to the energy dispersive X-ray spectroscopy (EDX) and Raman spectroscopy, and more importantly, the thickness is controllable. As shown in Fig. 10(e), the FET mobility of the InSe film is at the range of 10-70 $cm^2$/Vs with thickness ranging from 1nm to 20 nm, which is relatively good compared to other large-scale 2D materials. Photodetectors was fabricated by the InSe thin films, as displayed in Fig. 10(f). The maximal photoresponsivity is up to 27 A/W ($\lambda$ = 370 nm, P = 12.07 μW/$cm^2$ $V_{ds}$ = 1 V and $V_g$ = 0 V). The rise time and fall time are 0.5 s and 1.7 s, respectively. To sum up, the photodetector based on wafer-scale InSe thin films exhibits a relatively good performance, which hold promise in future applications.

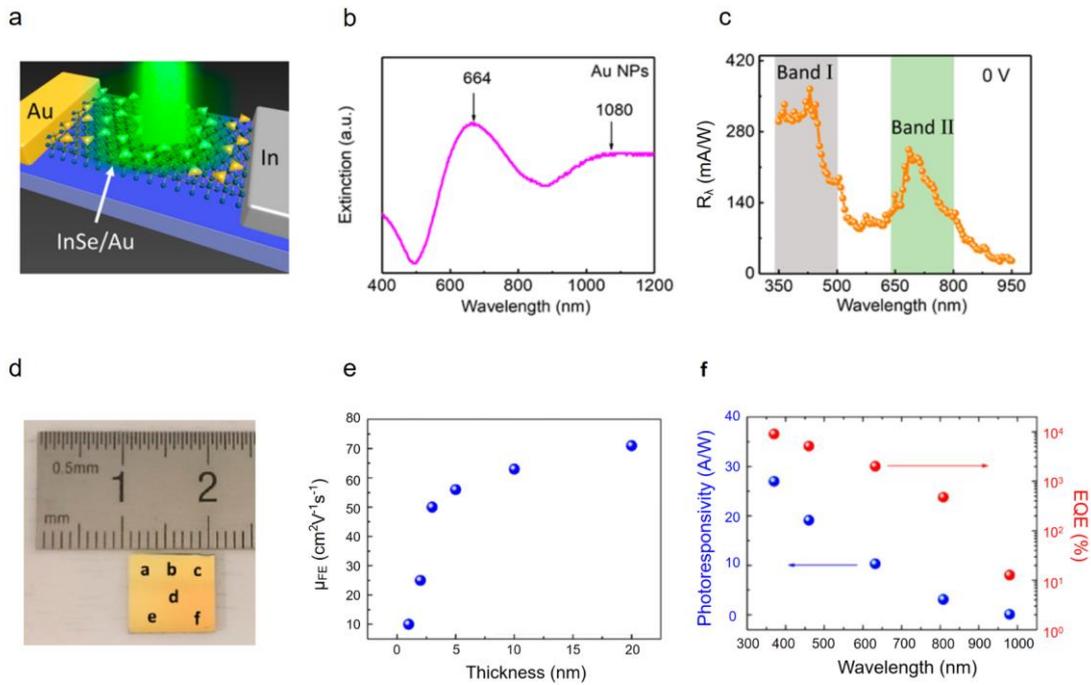

Fig. 10 (a) The schematic illustration of the Au plasmon/InSe photodetector. (b) The extinction spectrum of the Au NPs. (c) The wavelength-dependent photoresponsivity of the Au plasmon/InSe photodetector[107]. Reproduced with permission from ACS Nano 12, 8739 (2018). Copyright 2018 American Chemical Society. (d) The image of the centimeter-scale InSe thin film. (e) The FET mobility of the InSe thin films. (f) The wavelength-dependent photoresponsivity of the photodetector based on wafer-scale InSe thin films[105]. Reproduced with permission from ACS Nano 11, 4225 (2017). Copyright 2017 American Chemical Society.

### B. Heterostructures

InSe has been fabricated into vdW heterostructure photodetectors with other 2D materials such as GaSe[97, 104], GaTe[108, 117], BP[38, 112, 113], CuInSe$_2$[147] etc. GaSe is a p-type semiconductor with an indirect bandgap of about 2.0 eV[145]. As III-VI group layered materials, GaSe and InSe have similar crystal structures and close lattice parameters, which can be easily fabricated into heterostructures. Balakrishnan et al. fabricated a p-n junction with p-type GaSe and n-type InSe[97]. The InSe/GaSe heterostructure has type-II band alignment as illustrated in Fig. 11(a)[148]. The conduction band minimum (valence band maximum) of GaSe lies above that of InSe by 0.9 eV (0.1 eV). Electroluminescence (EL) originated from the indirect interlayer exciton at the InSe/GaSe interface was observed at room temperature. As shown in Fig. 11(b), the EL peak exhibits prominent redshift by about 0.1 eV compared to the PL peak of pristine InSe. Yan et al. designed a self-driven photodetector based on the InSe/GaSe heterostructure[104]. As demonstrated in Fig. 11(c), graphene was used as contact electrodes to build the vertical heterostructure. Owing to the type II band alignment and the built-in potential of the heterostructure, InSe/GaSe photodetector can operate at zero bias voltage with photoresponsivity up to 21 mAW$^{-1}$ ($\lambda$ = 410 nm). As displayed in Fig. 11(d), the photodetector exhibits multicolor photoresponse ranging from ultraviolet to near-infrared range (270-920nm). The maximal photoresponsivity is 350 AW$^{-1}$ ($\lambda$ = 410 nm, P = 0.025 mWcm$^{-2}$, $V_{ds}$ = 2V). Moreover, the response time of InSe/GaSe photodetector is extremely short. The rise time and fall time are 5.97 μs and 5.66 μs respectively at zero bias voltage ($\lambda$ = 470 nm) as shown in Fig. 11(e), which are prominently shorter than other vdW heterostructures. The vdW heterostructures with type-II band alignment can exhibit broader range photoresponse than the pristine materials due to the formation of interlayer exciton at the interface. Qi et al. fabricated a InSe/GaTe photodetector[117] as shown in Fig. 11(f). The photodetector can work at long wavelength up to 1.55 um (see Fig. 11(g)). In comparison, the optical band gap of few-layer GaTe is at 1.70 eV ($\lambda$ = 730 nm) and that of few-layer InSe is at 1.30 eV ($\lambda$ = 950 nm). DFT calculations demonstrated that the transition of interlayer exciton is approximately at 0.55 eV (see Fig. 11(h)), consistent with the photoresponse of InSe/GaTe heterostructures.

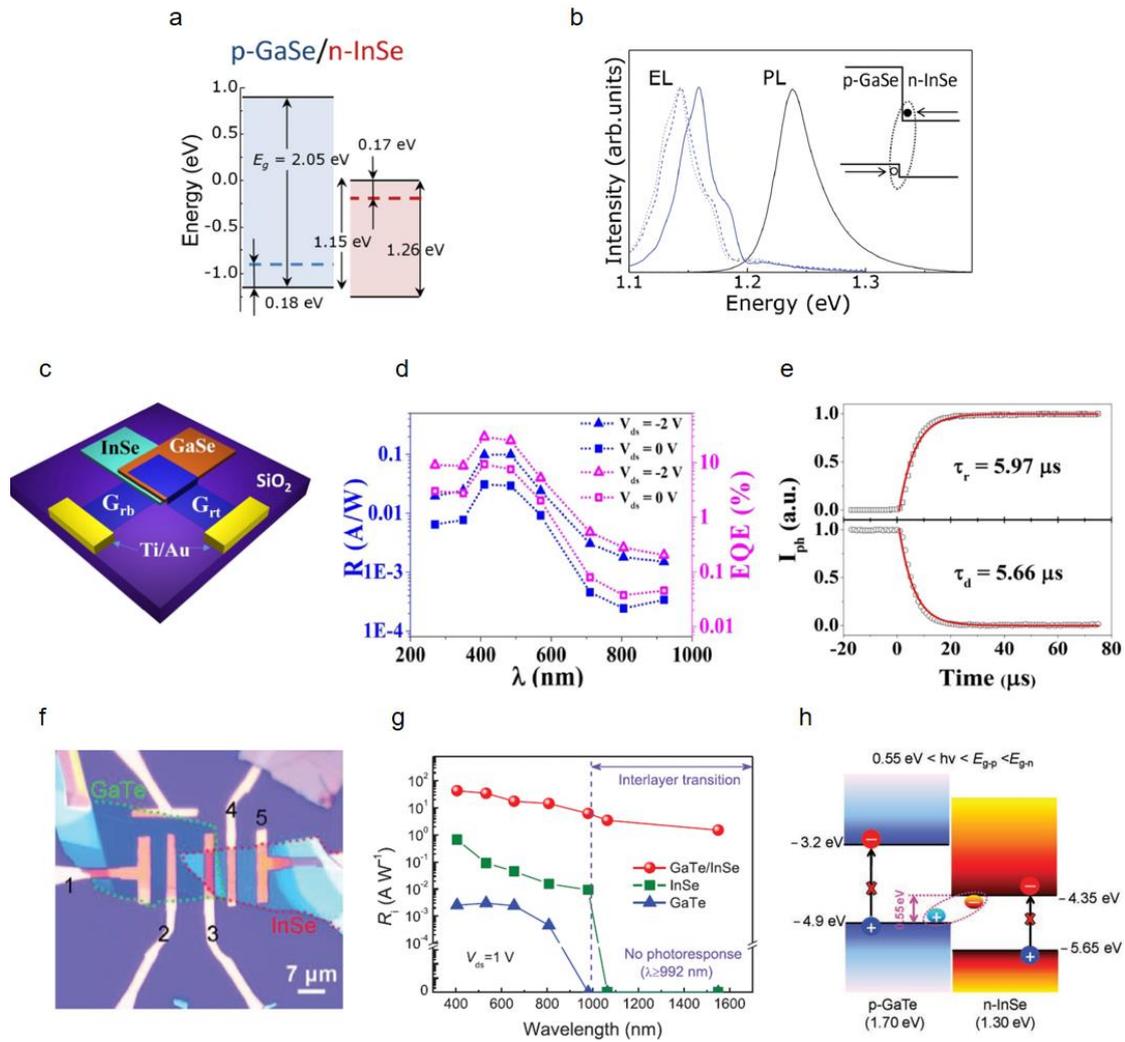

Fig. 11 (a) The band diagram of bulk InSe/GaSe heterostructures. (b) The EL and the PL spectra of InSe/GaSe heterostructures and thick InSe respectively[97]. Reproduced with permission from Adv. Optical Mater. 2, 1064 (2014). Copyright 2014 the Authors. Published by WILEY-VCH Verlag GmbH & Co. KGaA, Weinheim. (c) The schematic illustration of an InSe/GaSe photodetector. (d) The wavelength-dependent photoresponsivity (P = 1 mWcm$^{-2}$) and (e) the dynamic photoresponse of an InSe/GaSe photodetector[104]. Reproduced with permission from Nanotechnology 28 (2017). Copyright 2017 IOP Publishing Ltd. (f) The image and (g) the wavelength-dependent photoresponsivity of an InSe/GaTe photodetector. (h) The band diagram of bulk InSe/GaTe heterostructures[117]. Reproduced with permission from Adv. Funct. Mater. 30, 1905687 (2020).

InSe and BP have very similar physical properties, both possessing strongly layer-dependent direct band gap and high carrier mobility. It is of great interest to study the vdW heterostructures based on them. Theoretical calculations indicate that InSe/BP heterostructures belong to type-II band alignment[149-152]. Fig. 12(a) shows the DFT band structure and the band diagram for monolayer InSe/bilayer BP heterostructure[150]. The valence band maximum (conduction band minimum) is in BP (InSe) layers. The type-II band alignment persists for thick InSe/BP heterostructures. InSe/BP heterostructures have been utilized to fabricate high performance FETs and photodetectors[38, 95, 96, 112, 113]. Zhao et al. constructed a vertical p-n diode by stacking p-type BP on n-type InSe[38] as illustrated in Fig. 12(b). The heterostructure exhibits broadband photoresponse from the visible to near-infrared range (455-920 nm) as shown in Fig. 12(c). The maximal photoresponsivity is 11.7 mAW$^{-1}$ ($\lambda$ = 455 nm, P = 12.8 Wmm$^{-2}$, $V_{ds}$ = 0.5 V), which is comparable to that of BP/MoS$_2$ photodetector[153]. The photodetector shows fast photoresponse during the on-off switching process as shown in Fig. 12(d). The rise time and the fall time are estimated to be 24 ms and 32 ms respectively. In addition, due to the in-plane anisotropy of BP, the photodetector exhibits highly polarized photocurrent (see Fig. 12(e)). The anisotropy ratio is up to 0.83 (zero bias voltage, $\lambda$ = 633 nm), which is higher than that of prinstine BP devices (~0.3)[154]. Gao et al. fabricated avalanche photodetectors based on InSe/BP heterostructures[113]. As shown in Fig. 12(f), in dark environment, the reverse biased current suddenly increases by five orders of magnitude above a threshold voltage (about -4.8 V). This is a typical behavior of avalanche breakdown arising from impact ionization process. By taking advantage of avalanche breakdown, the photodetector exhibits extremely large multiplication factor (up to $3 \times 10^4$ on breakdown mode as displayed in Fig. 12(g)) and low noises under the illumination of a 4 μm laser. The small avalanche voltage, low noises and unconventional positive temperature coefficient indicate that the avalanche in InSe/BP heterostructures belongs to a new mode - ballistic avalanche. In ballistic avalanche, the ionizing collision happens only one time per primary carrier transit as illustrated in Fig. 12(h). Overall, the avalanche photodetector based on InSe/BP heterostructures is promising for weak light detection owing to the superior performance.

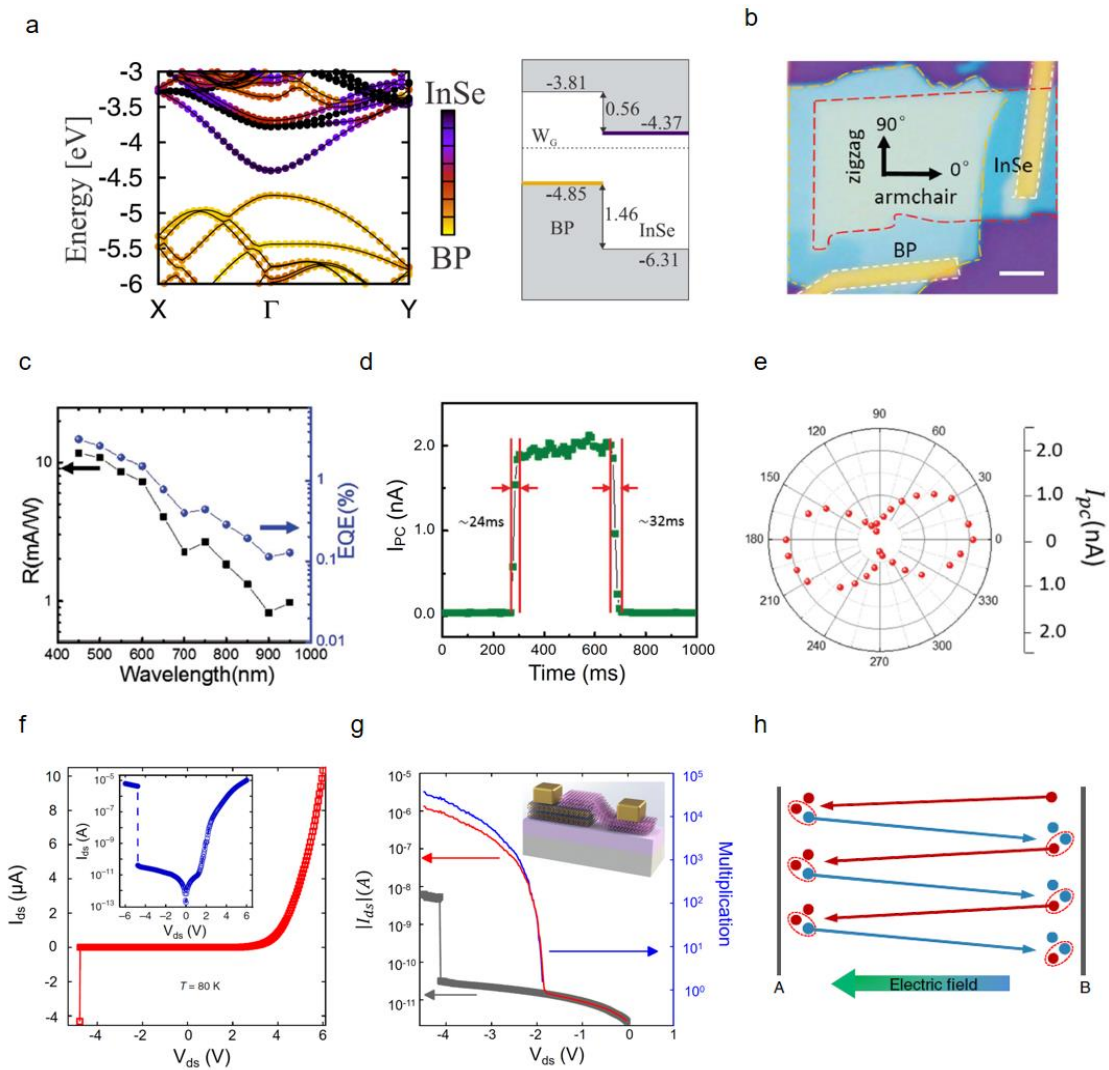

Fig. 12 (a) The DFT band structure and the band diagram of a monolayer InSe/bilayer BP

heterostructure[150]. Reproduced with permission from Phys. Rev. B 95, 195143 (2017). Copyright 2017 American Physical Society. (b) The image of an InSe/BP photodetector. (c) The wavelength-dependent photoresponsivity, (d) the dynamic photoresponse and (e) the polarized photocurrent of the InSe/BP heterostructure[38]. Reproduced with permission from Adv. Funct. Mater. 28, 1802011 (2018). Copyright 2018 WILEY-VCH Verlag GmbH & Co. KGaA, Weinheim. (f) The I-V characteristics and (g) the multiplication factor of InSe/BP avalanche devices. (h) Schematic illustration of the ballistic avalanche process[113]. Reproduced with permission from Nat. Nanotechnol. 14, 217 (2019). Copyright 2019, Springer Nature.

## V. Outlook and Conclusions

In this perspective, we systematically review the optical properties of few-layer InSe mainly from the experimental point of view. Few-layer InSe has been demonstrated as a promising 2D semiconductor with superior electronic properties and strongly layer-dependent band structures. To further discover intriguing optical properties of few-layer InSe and enhance the performance of its optoelectronics devices, how to effectively excite and detect the band gap transition is the first and foremost question we need to address. For instance, complex excitonic effects such as higher order states of free excitons, trions and biexcitons are expected to be observed. Although spin-orbital interactions can partially relax the transition selection rules[36, 45], the absorption of in-plane polarized light near the band gap transition is still extremely weak. Lamella-shaped structure can allow the illumination of out-of-plane polarized light to the samples[37, 39, 56]. It has already made progress in studying the optical transitions between valence subbands and conduction band of few-layer InSe[37]. Grazing incident light or coupling to planar waveguides can possibly play a role in future studies[155-

[158]. Secondly, as a natural quantum well, few-layer InSe exhibits potentials in optoelectronic applications in infrared and terahertz region by making use of the intersubband transitions[37]. For example, the intersubband transitions of $WSe_2$ quantum wells have been spectrally resolved by near-field spectroscopy[159]. Thirdly, the intrinsic optical and electronic properties of 2D materials can be manipulated by vdW heterostructures, which results in many fancy phenomena such as moiré excitons[160-162] and superconductivity[163]. The robust interlayer excitons at the vdW interface between InSe and TMDCs have been reported[39]. In the future, there is an abundance of opportunities in InSe heterostructures with other 2D materials. Particular attention should be paid to BP, which has in-plane anisotropy, layer-dependent direct band gap and pronounced excitonic effect[30-33]. Fourthly, few-layer InSe can be fabricated into or coupled to plasmonic nanostructures. The light-matter interactions can be largely enhanced or modified, which plays a role in a variety of applications such as excitonic lasers[164] and sensors[165]. Last but not least, the oxidation and degradation at ambient environment can dramatically influence the optical and electronic properties of few-layer InSe[35, 69, 86, 91, 93, 106, 166-175]. Hence it is imperative to find a way to prevent the oxidation and degradation. Encapsulation by hBN is a widely applied method. The PL spectrum and electron transport properties can remain unchanged for several weeks or even months[35, 91]. In addition, the ambient degradation can be suppressed by dry oxidation[86, 93, 173] or encapsulation with alumina films[175]. In summary, InSe is a competitive 2D material with unique optical and electronic properties, which possess great potentials not only in fundamental research, but also in the electronic and optoelectronic applications.

**Acknowledgments**

H.Y. is grateful to the financial support from the National Natural Science Foundation of China (Grant No. 11874009, 11734007), the National Key Research and Development Program of China (Grant No. 2016YFA0203900 and 2017YFA0303504) and Strategic Priority Research Program of Chinese Academy of Sciences (XDB30000000). C.W. is grateful to the financial support from the National Natural Science Foundation of China (Grant No. 11704075).

**Data Availability**

The data that support the findings of this study are available from the corresponding author upon reasonable request.